\documentclass[10pt,aps,pra,twocolumn,showpacs,superscriptaddress]{revtex4-2}

\usepackage[colorlinks=true,linkcolor=blue,citecolor=blue,urlcolor=magenta]{hyperref}

\usepackage{amsfonts,mathrsfs,amsmath,amsthm,amssymb}
\usepackage{epsfig}
\usepackage{bm}
\usepackage{array,multirow}
\usepackage{booktabs}
\usepackage{upgreek}
\usepackage{braket}
\usepackage[utf8]{inputenc}
\usepackage{ulem}

\begin{document}  
\title {\bf 
Exact gradient for general cost functions in variational quantum algorithms}

\author{Jesus Urbaneja} 
\affiliation{Graduate School of Information Science, Tohoku University, Sendai 980-0845, Japan}
\affiliation{Frontier Research Institute 
for Interdisciplinary Sciences, 
Tohoku University, Sendai 980-8578, Japan}

\author{Le Bin Ho} 
\thanks{Electronic address: binho@fris.tohoku.ac.jp}
\affiliation{Frontier Research Institute 
for Interdisciplinary Sciences, 
Tohoku University, Sendai 980-8578, Japan}
\affiliation{Department of Applied Physics, 
Graduate School of Engineering, 
Tohoku University, 
Sendai 980-8579, Japan}

\date{\today}

\begin{abstract}
We present a unitary-based gradient formulation for variational quantum algorithms (VQAs) that applies to general differentiable cost function defined by a parameterized quantum circuit composed of Pauli-generated rotations. The gradient is obtained directly from the underlying unitary evolution, without assuming a specific expectation-value form of the cost function. The resulting expressions can be accessed on quantum hardware using the Hadamard and Hilbert-Schmidt tests. We demonstrate the method in variational quantum compilation, where it yields stable and accurate gradient estimates. This unitary-based framework therefore provides a broadly applicable and hardware-compatible tool for gradient evaluation in VQAs.
\end{abstract}
\maketitle

\sloppy

\section{Introduction}
Variational quantum algorithms (VQAs) are a central approach for utilizing noisy intermediate-scale quantum (NISQ) devices \cite{Cerezo2021}. They use a hybrid quantum-classical workflow where a quantum processor prepares a parameterized circuit and a classical optimizer updates the parameters to minimize a chosen cost function. VQAs have been applied to several areas, including quantum chemistry via the variational quantum eigensolver (VQE) \cite{Peruzzo2014}, combinatorial optimization through the quantum approximate optimization algorithm (QAOA) \cite{farhi2014quantumapproximateoptimizationalgorithm,BLEKOS20241}, quantum machine learning \cite{Biamonte2017}, and variational quantum compilation (VQC) \cite{TuanHai_2024}.

The cost function lies at the center of VQAs and is defined using a parameterized quantum circuit \(\mathcal{U}(\boldsymbol{\theta})\) with parameters \(\boldsymbol{\theta}=(\theta_1,\theta_2,\ldots)\). Efficient optimization requires accurate and hardware-compatible evaluation of its gradients with respect to these parameters. A widely used approach is the parameter-shift rule (PSR), which computes derivatives by evaluating the cost function at shifted values of the parameters \cite{PhysRevA.98.032309,PhysRevA.99.032331}.
However, the standard PSR is formulated at the expectation-value cost-function level and relies on restrictive assumptions, such as single-qubit Pauli generators or expectation values with specific algebraic structures. These assumptions define the scope in which the rule is directly applicable. As a result, its applicability is limited for more general cost functions or for circuits whose generators have richer spectra. 
Related expectation-based and operator-level gradient formulations have also been developed \cite{PhysRevA.98.032309,crooks2019}, providing general tools for analytic gradient evaluation within expectation-value-based frameworks.
Extensions such as the stochastic PSR \cite{Banchi2021measuringanalytic,Ho2023}, photonic PSR \cite{PhysRevA.111.032429}, and generalized PSR frameworks \cite{Wierichs2022generalparameter,PhysRevA.104.062443,Hai2024} relax some of these gate-local constraints, but still remain formulated in terms of expectation-value cost functions and do not provide a fully general rule for arbitrary differentiable costs.

In this work, we present a unitary-based and exact formulation of the PSR that differs from existing approaches: instead of operating at the cost-function (expectation-value) level, we derive the shift rule directly from the parameterized unitary operator itself. This unitary-centric formulation enables gradient evaluation for any differentiable cost function that depends on the circuit unitary, without assuming an expectation-value structure. By differentiating the unitary and using the structure of Pauli-generated rotations, we obtain a unitary-based PSR in which each derivative reduces to evaluating the circuit at a \(\pi\)-shifted parameter. At the unitary level, the gradient expression holds for arbitrary differentiable cost function that depends on \(\mathcal{U}(\boldsymbol{\theta})\), regardless of its internal form. This clearly distinguishes our method from the standard expectation-based PSR and provides a hardware-compatible gradient framework.

We illustrate the approach using a VQC model  where a parameterized unitary \(\mathcal{U}(\boldsymbol{\theta})\) is trained to approximate a target unitary \(\mathcal{V}\) up to a global phase. Using the Hilbert-Schmidt inner product as the cost function, we show that the gradient obtained from the unitary-based PSR enables efficient and stable optimization. The gradient can be measured directly on quantum hardware via the Hadamard test (HT) and the Hilbert-Schmidt test (HST), providing a practical route for implementation.

As a case study, we examine the VQC of a two-qubit Greenberger-Horne-Zeilinger (GHZ) state using an ansatz composed of \(R_y, R_z\), and CNOT gates. We benchmark the unitary-based PSR against numerical finite-difference methods, including two-point and four-point formulas. Our approach achieves exact accuracy and remains stable even at small step sizes, where finite differences become highly sensitive to numerical noise, while remaining compatible with near-term quantum devices. We then extend the analysis to the synthesis of \(N\)-qubit Toffoli gates and Haar random gates, where the unitary-based PSR reproduces the theoretical gradient, whereas the standard PSR shows systematic deviations from the expected result.

This work, thus, presents a unitary-based formulation of the PSR that applies to arbitrary VQA cost functions. The approach offers a clear gradient expression and practical benefits for gradient-based optimization.

The paper is organized as follows. Section~\ref{spsr} reviews the limitations of the standard PSR.
In Sec.~\ref{sec2}, we derive the unitary-based exact gradient formula for general VQA models.
Section~\ref{sec3} applies this framework to variational quantum compilation, and Sec.~\ref{sec4} presents numerical demonstrations of the results.
A broader discussion is provided in Sec.~\ref{sec5}, and Sec.~\ref{sec6} concludes with a summary and outlook.

\section{Limitation of the standard PSR}
\label{spsr}
In this section, we examine the limitations of the standard PSR. Although it is often regarded as a ``universal" analytic-gradient method for cost functions generated by parametrized unitaries with Pauli-type generators \cite{PhysRevA.98.032309,PhysRevA.99.032331,Wierichs2022generalparameter}, we show through a simple polynomial model that the PSR is exact only for a restricted class of functions.

For a real parameter \(x\), 
the standard PSR defines the derivative of a cost function \(f(x)\) by
\begin{equation}
f'(x) = r\big[f(x+s) - f(x-s)\big],
\label{eq:psr_scalar}
\end{equation}
where \(s\) is an arbitrary shift and \(r\) is a constant prefactor
(\(r = 1/(2s)\) in the central-difference form, or \(r = 1/(2\sin s)\) in the
usual trigonometric formulation).  We now test Eq.~\eqref{eq:psr_scalar} on
polynomials
\begin{equation}
f(x) = P_n(x) = a_n x^n + \cdots + a_1 x + a_0,
\end{equation}
and compare the result with the theory derivative \(f'(x)\).

\subsection{Linear and quadratic functions}
For a linear function \(f(x)=ax+b\), the exact derivative is \(f'(x)=a\). Substituting \(f(x\pm s)\) into Eq.~\eqref{eq:psr_scalar} with $r = 1/2s$ immediately gives
\begin{align}
    f'(x)=\frac{1}{2s}[f(x+s)-f(x-s)]=a,
\end{align}
so, the PSR holds for any nonzero \(s\).
For a quadratic function \(f(x)=ax^{2}+bx+c\), whose derivative is \(f'(x)=2ax+b\), expanding \(f(x\pm s)\), we find
\begin{align}
    f(x+s)-f(x-s)=2s(2ax+b),
\end{align}
and therefore
\begin{align}
    f'(x)=\frac{1}{2s}\big[f(x+s)-f(x-s)\big]=2ax+b.
\end{align}
Thus, the standard PSR is exact for both linear and quadratic polynomials.

\subsection{Cubic and higher-order functions}
For a cubic polynomial
\(
f(x) = a x^3 + b x^2 + c x + d,
\)
the true derivative is \(f'(x) = 3 a x^2 + 2 b x + c\).  Expanding \(f(x\pm s)\) gives
\begin{align}
f(x+s) - f(x-s)
&= 2 s \big( 3 a x^2 + 2 b x + c + a s^2 \big),
\end{align}
and therefore
\begin{equation}
\frac{1}{2s}\big[f(x+s) - f(x-s)\big]
= 3 a x^2 + 2 b x + c + a s^2.
\end{equation}
The extra term \(a s^2\) cannot be eliminated for any fixed \(s \neq 0\), so the standard PSR fails to reproduce the correct derivative. The same issue arises for all higher-order polynomials, where mixed terms \(x^k s^m\) with \(m \ge 2\)
appear in \(f(x+s) - f(x-s)\), preventing Eq.~\eqref{eq:psr_scalar} from matching \(f'(x)\) for all \(x\).

To obtain the correct derivative, one must effectively start from ``scratch" by decomposing the higher-order polynomial into lower-order components. For example,
\begin{equation}
f(x) = a x^{3} + b x^{2} + c x + d
     = x \,\big(a x^{2} + b x + c\big) + d.
\end{equation}
The linear and quadratic parts can then be handled separately using the standard PSR and a chain rule used to combine the results. Thus, the cubic case cannot be treated by a single application of the standard PSR; it must be reduced to simpler pieces for which analytic rules apply. This decomposition strategy extends naturally to all higher-order functions.

\subsection{Implications for VQAs}
The standard PSR is exact only when the cost function is a single expectation value, whose dependence on the parameters has a simple two-frequency structure set by the eigenvalues of the generator. For example, for Pauli-generated parameterized gates, the associated expectation values involve only two eigenvalues and can be decomposed into linear trigonometric polynomials, for which the standard PSR applies~\cite{Hai2024}. This spectral simplicity makes it possible to recover the derivative from two shifted evaluations. In more general settings, expectation values can form higher-order polynomial combinations, in which case generalized PSR formulations are required~\cite{Wierichs2022generalparameter,PhysRevA.104.062443,Hai2024}.

In practice, many VQA cost functions are not single expectations but composite quantities constructed from overlaps, fidelities, Hilbert-Schmidt norms, or other nonlinear combinations. Examples include \(|\langle\phi|\psi(\bm\theta)\rangle|^{2}\), the Hilbert-Schmidt overlap \(|\mathrm{tr}(V^{\dagger}U(\bm\theta))|^{2}\), and loss functions used in quantum machine learning. These costs break the harmonic structure, and the standard PSR no longer returns the correct gradient. In this work, we address this limitation by deriving the gradient directly from the full operator structure of \(\mathcal{U}(\boldsymbol{\theta})\), yielding a unitary-based PSR that remains valid for any differentiable cost functional.

\section{Unitary-based PSR for general VQA models}\label{sec2}
%\subsection{Model}
VQAs are hybrid quantum algorithms that combine quantum circuits with classical optimization. They use a parameterized quantum circuit, written as $\mathcal{U}(\bm{\theta})$, where $\bm{\theta}$ represents a set of tunable parameters. A classical computer adjusts these parameters iteratively to minimize a cost function specific to the problem. This optimization process steers $\mathcal{U}(\bm{\theta})$ toward generating quantum states or operations that closely approximate the target solution.

Typically, the unitary \( \mathcal{U}(\bm{\theta}) \) is defined based on a sequence of quantum gates 
\begin{align}\label{eq:u}
\mathcal{U}(\bm{\theta}) = U_M(\theta_M) \cdots U_2(\theta_2)U_1(\theta_1),
\end{align}
where the parameters \( \bm \theta = (\theta_1, \cdots, \theta_M) \) are the rotation angles of the quantum circuit, and $M$ is the number of parameters. Without loss of generality, we assume that
each unitary is defined by $U_j(\theta_j) = {\rm exp}(-i\theta_j P_j/2)$,
where $P_j$ is a Pauli matrix. This decomposition is the standard setting of VQAs constructed from Pauli-generated rotation gates. If the same parameter appears in multiple gates, the derivative with respect to that parameter is obtained by summing the contributions from each occurrence.

\subsection{Cost function}
The cost function can be expressed in a general and task-agnostic form as \cite{Cerezo2021}
\begin{align}
    C(\bm{\theta}) = f\Big[\mathcal{U}^\dagger(\bm{\theta}), \mathcal{U}(\bm{\theta}), \{\rho_k\}, \{O_k\}\Big]
\end{align} 
for a functional $f$, where $\{\rho_k\}$ is a set of input quantum states (e.g., from a training set), and $\{O_k\}$ is a set of Hermitian observables or measurement operators. This formula is flexible enough to encompass a wide variety of VQAs, including VQE, QAOA, and quantum machine learning models, by appropriately choosing the states, observables, and weighting scheme.

The partial derivative of the cost function with respect to parameter $\theta_j$ can be expressed as
\begin{align}\label{eq:df}
\dfrac{\partial C(\bm{\theta})}{\partial \theta_j}
= f\Big[
\mathcal{U}^\dagger(\bm{\theta})
\dfrac{\partial \mathcal{U}(\bm{\theta})}{\partial \theta_j}
+
\dfrac{\partial \mathcal{U}^\dagger(\bm{\theta})}{\partial \theta_j}\mathcal{U}(\bm{\theta}), \cdots
\Big],
\end{align}
where $\cdots$ denotes $\{\rho_k\}$ and $\{O_k\}$ for brevity. The derivative terms are analytically tractable given the explicit form of $\mathcal{U}(\bm{\theta})$ defined in Eq.~\eqref{eq:u}. In the following, we derive the exact expression for Eq.~\eqref{eq:df}.

\subsection{Unitary-based PSR}
We first derive $\partial \mathcal{U}(\bm{\theta})/\partial \theta_j$ using \( \mathcal{U}(\bm{\theta}) \) in Eq.~\eqref{eq:u}, which gives
\begin{align}\label{eq:du}
\notag \dfrac{\partial \mathcal{U}(\bm{\theta})}{\partial \theta_j}
     &= U_M(\theta_M) \cdots 
     \dfrac{\partial U_j(\theta_j)}{\partial \theta_j}
     \cdots
     U_2(\theta_2)U_1(\theta_1)\\
     &= -\frac{i}{2} U_M(\theta_M) \cdots 
        P_jU_j(\theta_j)
        \cdots
        U_2(\theta_2)U_1(\theta_1),
\end{align}
where we applied $U_j(\theta_j) = {\rm exp}(-i\theta_j P_j/2)$.
Notable, when parameters enter nonlinearly in the exponent, such as \(U_j(\theta)=\exp[-if(\theta)P_j/2]\), the gradient acquires an additional prefactor \(f'(\theta)\) via the chain rule, while the unitary-based shift structure remains unchanged.

Given that $P_j$ is a Pauli matrix, i.e., $P_j^2 = I$, we decompose the unitary operator \( U_j(\theta_j) \) into
\begin{align}\label{eq:ude}
    U_j(\theta_j) = \cos\Big(\frac{\theta_j}{2}\Big) I -i\sin\Big(\frac{\theta_j}{2}\Big)P_j.
\end{align}
Now, by choosing \( \theta_j = \pi \), we find
$U_j(\pi) = -iP_j$.
Substituting this into Eq.~\eqref{eq:du}, we have
\begin{align}\label{eq:du1}
\notag \dfrac{\partial \mathcal{U}(\bm{\theta})}{\partial \theta_j} &= \frac{1}{2} U_M(\theta_M) \cdots 
        U_j(\theta_j+\pi)
        \cdots
        U_2(\theta_2)U_1(\theta_1)\\
        &= \frac{1}{2} \mathcal{U}(\bm\theta + \pi\bm e_j),
\end{align}
where \( \bm{e}_j \) is the unit vector in the \( j^{\text{th}} \) parameter, defined as \( (0, 0, \ldots, 1, \ldots, 0) \). Similarly, we have 
\begin{align}\label{eq:du1dag}
    \dfrac{\partial \mathcal{U}^\dagger(\bm{\theta})}{\partial \theta_j} = -\frac{1}{2} \mathcal{U}^\dagger(\bm\theta - \pi\bm e_j).
\end{align}

\begin{figure}[t!]
    \centering
\includegraphics[width=\columnwidth]{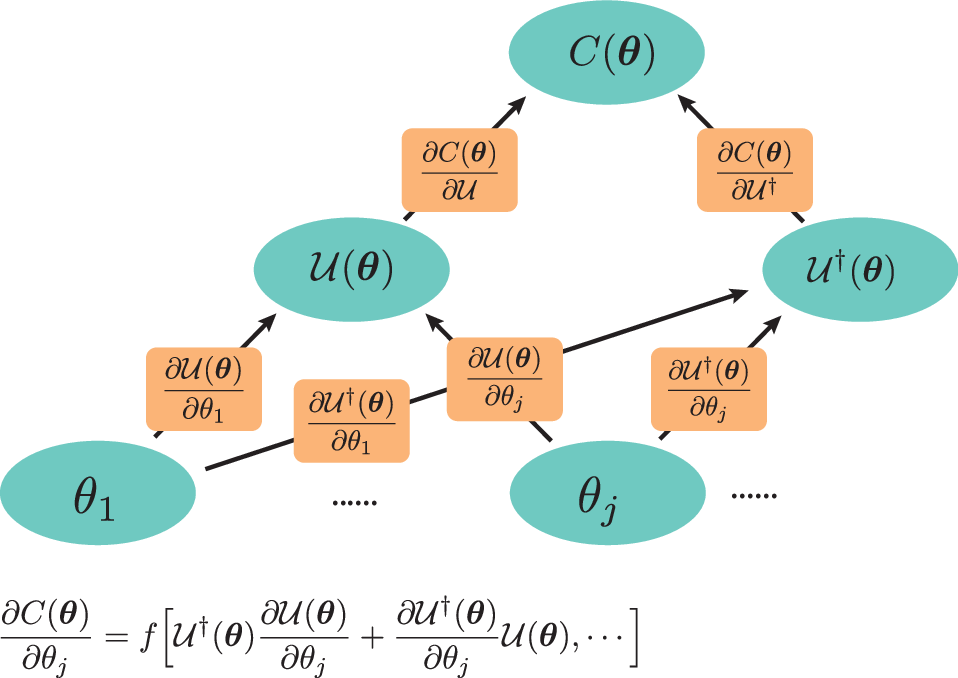}
    \caption{Schematic representation of the computational graph employed to compute the exact gradient of the cost function.
    }
    \label{fig1}
\end{figure}

These results show that the derivative $\partial \mathcal{U}(\bm{\theta})/{\partial \theta_j}$ and its Hermitian conjugate are directly accessible by shifting the gate $U_j(\theta_j)$ by an angle $\pi$ within the quantum circuit. Consequently, the gradient $\partial \mathcal{C}(\bm{\theta}) / \partial \theta_j$ can be naturally computed using these results.

Unlike the standard PSR, which applies a parameter-shift rule directly at the level of the cost function, our approach derives the gradient by differentiating the parameterized unitary itself. The shift rule is therefore constructed at the unitary level rather than in its original cost-function form. As a result, the gradient expression applies to any cost functional that depends differentially on \(\mathcal{U}(\boldsymbol{\theta})\), independent of how the cost is constructed from the circuit output.

Figure~\ref{fig1} shows the computational graph for evaluating the gradient of  $\mathcal{C}(\boldsymbol{\theta})$. The graph tracks how the parameters enter through the parameterized circuit $\mathcal{U}(\boldsymbol{\theta})$ and its Hermitian conjugate $\mathcal{U}^\dagger(\boldsymbol{\theta})$. Gradients $\partial \mathcal{C}(\boldsymbol{\theta}) / \partial \theta_j$ are obtained by back-propagating through this structure using the chain rules.

\subsection{Example for VQE}
We consider a VQE algorithm as an example. The goal is to approximate the ground state of a given Hamiltonian $H$ by minimizing its expectation value over a parameterized quantum state. VQEs have been successfully applied to estimate ground-state energies in quantum chemistry~\cite{Peruzzo2014,Lim2024},
simulate strongly correlated materials \cite{PhysRevApplied.23.044028},
and explore molecular systems on real quantum hardware \cite{ghavami2025groundstateenergymolecule,PhysRevA.111.022817,Guo2024}, making them a versatile tool across chemistry, physics, and materials science.
In VQEs, the cost function is defined by
\begin{align}
    \mathcal{C}(\boldsymbol{\theta}) = \langle \psi(\boldsymbol{\theta}) | H | \psi(\boldsymbol{\theta}) \rangle,
\end{align}
where $| \psi(\boldsymbol{\theta}) \rangle = \mathcal{U}(\bm\theta)|\psi_0\rangle$ is the variational ansatz prepared by acting $\mathcal{U}(\bm\theta)$ on an initial state $|\psi_0\rangle$.
We have 
\begin{align}
    \mathcal{C}(\boldsymbol{\theta})  %\equiv %f\big[\mathcal{U}(\bm\theta)\big] 
    = \langle\psi_0|\mathcal{U}^\dagger(\bm\theta)H\mathcal{U}(\bm\theta)|\psi_0\rangle,
\end{align} 
where its derivative gives
\begin{align}\label{eq:derC}
\frac{\partial \mathcal{C}(\boldsymbol{\theta})}{\partial \theta_j} = \bra{\psi_0}\Big[\mathcal{U}^\dagger H \big(\partial_{\theta_j} \mathcal{U}\big)+\big(\partial_{\theta_j} \mathcal{U}^\dagger \big) H\mathcal{U}\Big]\ket{\psi_0},
\end{align} 
where $\mathcal{U}\equiv\mathcal{U}(\bm\theta)$ for short.
Substituting Eqs.~(\ref{eq:du1}, \ref{eq:du1dag}) into Eq.~\eqref{eq:derC}, we obtain
\begin{align}
    \notag\frac{\partial \mathcal{C}(\boldsymbol{\theta})}{\partial \theta_j} 
    & = \frac{1}{2}\langle\psi_0|\Big[\mathcal{U}^\dagger H \mathcal{U}(\bm\theta+\pi\bm e_j)-\mathcal{U}^\dagger(\bm\theta-\pi\bm e_j) H\mathcal{U}\Big]|\psi_0\rangle\\
    &=\frac{1}{2}\langle\psi_0|\mathcal{U}^\dagger
    \Big[H \mathcal{U}(\pi\bm e_j)
    -\mathcal{U}^\dagger(-\pi\bm e_j) H\Big]\mathcal{U}|\psi_0\rangle.
\end{align}
Then, using the relations
\begin{align}
\notag H\, \mathcal{U}(\pi \bm{e}_j) &= U^\dagger\left(\tfrac{\pi}{2} \bm{e}_j\right) \, H \, \mathcal{U}\left(\tfrac{\pi}{2} \bm{e}_j\right), \\
\notag \mathcal{U}^\dagger(-\pi\bm e_j) H &= \mathcal{U}^\dagger(-\tfrac{\pi}{2}\bm e_j) \, H \, \mathcal{U}(-\tfrac{\pi}{2}\bm e_j),
\end{align}
we obtain the exact derivative with the form
\begin{align}
    \notag\frac{\partial \mathcal{C}(\boldsymbol{\theta})}{\partial \theta_j} 
    & = \frac{1}{2}\langle\psi_0|\Big[\mathcal{U}^\dagger(\bm\theta + \tfrac{\pi}{2} \boldsymbol{e}_j) H \mathcal{U}(\bm\theta + \tfrac{\pi}{2} \boldsymbol{e}_j) \\
    &\hspace{1cm} -\mathcal{U}^\dagger(\bm\theta - \tfrac{\pi}{2} \boldsymbol{e}_j) H\mathcal{U}(\bm\theta - \tfrac{\pi}{2} \boldsymbol{e}_j)\Big]|\psi_0\rangle,
\end{align}
which recasts to
\begin{align}
    \frac{\partial \mathcal{C}(\boldsymbol{\theta})}{\partial \theta_j} 
    = \frac{\mathcal{C}(\boldsymbol{\theta} + \frac{\pi}{2} \boldsymbol{e}_j) - \mathcal{C}(\boldsymbol{\theta} - \frac{\pi}{2} \boldsymbol{e}_j)}{2}.
\end{align}
This is a well-established PSR for computing gradients in VQEs~\cite{Wierichs2022}.
To compute the derivative $\partial \mathcal{C}(\boldsymbol{\theta}) / \partial \theta_j$, we measure the cost function twice: once with the parameter shifted to $\boldsymbol{\theta} + \frac{\pi}{2} \boldsymbol{e}_j$ and once with $\boldsymbol{\theta} - \frac{\pi}{2} \boldsymbol{e}_j$. The derivative is then obtained by subtracting the two results.

% \subsection{Tree}
\section{Application to VQC}\label{sec3}
To demonstrate our method, we consider a VQC framework, where a parameterized unitary $\mathcal{U}(\boldsymbol{\theta})$ is optimized to approximate a target unitary $\mathcal{V}$ \cite{Monotten2004,TuanHai_2024}. The objective is to train $\mathcal{U}(\boldsymbol{\theta})$ so that it reproduces the action of $\mathcal{V}$ up to a global phase, satisfying
\begin{align}\label{eq:VU}
    \mathcal{V}^\dagger \mathcal{U}(\bm{\theta}) = e^{-\phi} I,
\end{align}
where $\phi \in \mathbb{R}$ denotes a global phase \cite{Khatri2019quantumassisted}. This condition ensures that $\mathcal{U}(\bm{\theta})$ faithfully reproduces the target unitary transformation. This approach has been used in gate optimization \cite{heya2018variational}, quantum-assisted compiling \cite{Khatri2019quantumassisted}, quantum state tomography \cite{hai2023universal}, and quantum object simulations \cite{TuanHai_2024,HAI2024101726}.

%\subsection{Cost function}
To optimize the scheme, we use the Hilbert-Schmidt test as the cost function
\begin{align}\label{eq:cost}
    \mathcal{C(\bm\theta}) = 1-\frac{1}{d^2} \Big| \text{tr} \big[\mathcal{V}^\dagger \mathcal{U} (\bm \theta) \big]\Big|^2,
\end{align}
where \( d \) is the dimension of the Hilbert space. 
When $\mathcal{C}(\bm{\theta}) = 0$, Eq.\eqref{eq:VU} is satisfied.
This cost function is nonlinear, and therefore the standard PSR cannot be applied directly in its original form. Instead, the gradient must be obtained by differentiating the cost function as a composite function of the circuit parameters and then combining the resulting derivative terms \cite{Khatri2019quantumassisted}.

\subsection{Unitary-based PSR solution}
To optimize \( \bm{\theta} \), we need to compute the derivatives of the cost function with respect to these parameters. Here, we apply the unitary-based PSR solution. First, we derive
\begin{align}
    \notag \frac{\partial \mathcal{C(\bm\theta})}{\partial \theta_j} 
    &= -\frac{1}{d^2} \frac{\partial \left| \text{tr} \left( \mathcal{V}^\dagger \mathcal{U} \right) \right|^2}{\partial \theta_j} \\
    &= -\frac{1}{d^2} \Bigg[ 
        \text{tr} \left( \frac{\partial \left(\mathcal{V}^\dagger \mathcal{U} \right)}{\partial \theta_j} \right)
        \text{tr} \left( \mathcal{V}^\dagger \mathcal{U} \right)^* \nonumber \\
    &\hspace{4em} + \text{tr} \left( \mathcal{V}^\dagger \mathcal{U} \right)
        \text{tr} \left( \frac{\partial \left(\mathcal{V}^\dagger \mathcal{U}\right)}{\partial \theta_j} \right)^*
    \Bigg].
    \label{eq:theory_prev}
\end{align}
Then, we obtain
\begin{align}
       \frac{\partial \mathcal{C(\bm\theta})}{\partial \theta_j} = -\frac{2}{d^2} \text{Re} \Bigg[ \text{tr} \left( \frac{\partial \left(\mathcal{V}^\dagger \mathcal{U}\right)}{\partial \theta_j} \right) \text{tr}(\mathcal{V}^\dagger \mathcal{U})^* \Bigg].
    \label{eq:exact}
\end{align}
Using the unitary-based PSR solution Eq.~\eqref{eq:du1}, we get
\begin{align}
\frac{\partial \mathcal{C(\bm\theta})}{\partial \theta_j} = -\frac{1}{d^2} \text{Re} \big[ \text{tr} \left( \mathcal{V}^\dagger \mathcal{U}(\bm\theta+\pi\bm{e}_j) \right) \text{tr} \left( \mathcal{V}^\dagger \mathcal{U}(\bm\theta)\right)^{*}\big].
\label{eq:theory}
\end{align}

To implement Eq.~\eqref{eq:theory} on quantum circuits, we consider two distinct quantum subroutines: the Hadamard test (HT) and the Hilbert-Schmidt test (HST), then we compare their performance.

\subsection{Quantum-circuit-based implementation}

In this subsection, we show how to evaluate Eq.~\eqref{eq:theory} via calculating \(\mathrm{tr}(\mathcal{V}^\dagger\mathcal{U})\) using HT and HST methods. Both of them provide scalable means to access overlaps between unitary circuits, which are central to our cost function and gradient evaluations.

\begin{figure}
    \centering 
\includegraphics[width=\columnwidth]{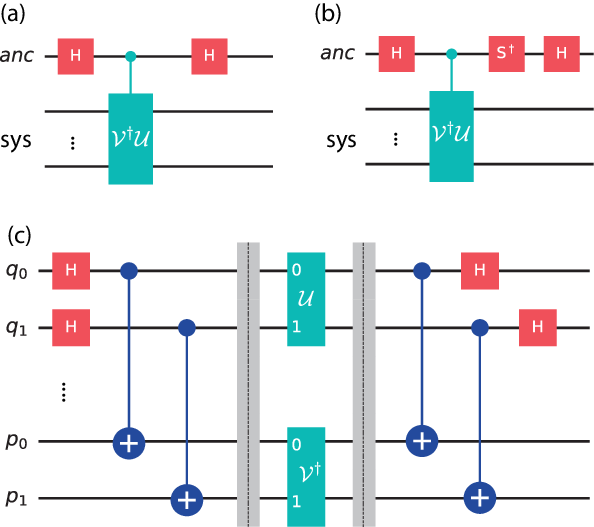} 
    \caption{Quantum circuits for HT (a,b) and HST (c).}
    \label{fig2} 
\end{figure}

\subsubsection{Hadamard test}
We begin by expressing the trace \(\mathrm{tr}(\mathcal{V}^\dagger \mathcal{U}) \) in Eq.~\eqref{eq:theory} as a sum over computational basis states
\begin{align}
\mathrm{tr}(\mathcal{V}^\dagger \mathcal{U}) = \sum_{j=0}^{2^n - 1} \bra{j} \mathcal{V}^\dagger \mathcal{U} \ket{j},
\end{align}
where $n$ is the number of qubits, i.e., the dimension is $d = 2^n$.
Each term $\bra{j} \mathcal{V}^\dagger \mathcal{U} \ket{j}$ can be calculated via the Hadamard test: An ancilla is initialized in $\ket{0}$ and the system prepared in $\ket{j}$; a Hadamard gate is applied to the ancilla, followed by a controlled-$\mathcal{V}^\dagger \mathcal{U}$ with the ancilla as control; a second Hadamard is then applied to the ancilla before measurement in the computational basis. See details in Fig.~\ref{fig2}(a). The resulting expectation value satisfies
\begin{align}
\langle \sigma_z \rangle^{\rm R}_j = \text{Re}[\bra{j} \mathcal{V}^\dagger \mathcal{U} \ket{j}],
\end{align}
where the subscript R stands for `Real'. To obtain the imaginary part, a phase gate $S = \mathrm{diag}(1, i)$ is inserted before the second Hadamard [Fig.~\ref{fig2}(b)], yielding
\begin{align}
\langle \sigma_z \rangle^{\rm I}_j = \text{Im}[\bra{j} \mathcal{V}^\dagger \mathcal{U} \ket{j}].
\end{align}

By averaging over all $2^n$ computational basis states (or sampling a subset for efficiency), we estimate
\begin{align}
\mathrm{tr}(\mathcal{V}^\dagger \mathcal{U}) = \sum_j \Big( \langle \sigma_z \rangle^{\rm R}_j + i\, \langle \sigma_z \rangle^{\rm I}_j \Big).
\end{align}
For larger $n$, sampling from a 1-design ensemble (e.g., Haar-random or Pauli-basis states) and averaging,
\begin{align}
\mathrm{tr}(\mathcal{V}^\dagger \mathcal{U}) = 2^n\ \mathbb{E}_{\ket{\psi}} \left[ \bra{\psi} \mathcal{V}^\dagger \mathcal{U} \ket{\psi} \right],
\end{align}
yields an unbiased estimator of the trace, which is often more practical for large Hilbert spaces.

Similarly, we derive \(\mathrm{tr}\big(\mathcal{V}^\dagger \mathcal{U}(\bm\theta+\pi\bm{e}_j)\big) \) in Eq.~\eqref{eq:theory} by shifting \(\bm\theta\) by an angle \(\pi/2\) and following the same procedure. Finally, substituting these results into Eq.~\eqref{eq:theory} yields the final implementation.

\subsubsection{Hilbert-Schmidt test}
The HST evaluates the trace overlap using the identity
\begin{align}
\mathrm{tr}(\mathcal{V}^\dagger \mathcal{U}) = 2^n \bra{\Phi^+}, \mathcal{V}^\dagger \otimes \mathcal{U}, \ket{\Phi^+},
\end{align}
where $\ket{\Phi^+} = 2^{-n/2} \sum_{i=0}^{2^n-1}\ket{i}\otimes\ket{i}$, 
a maximally entangled state of two 
$n$-qubit registers.

The scheme prepares entanglement by applying Hadamard gates to the first register, followed by CNOTs to the second. The unitaries $\mathcal{V}^\dagger$ and $\mathcal{U}$ are then applied to the two registers, and the entangling operations are reversed to complete the protocol [Fig.~\ref{fig2}(c)].

The probability of obtaining the all-zero outcome when measuring $2n$ qubits is
\begin{align}
P_0 = \frac{1}{2^{2n}} \left|\mathrm{tr}(\mathcal{V}^\dagger \mathcal{U})\right|^2 .
\end{align}
Using this expression, the cost function Eq.~\eqref{eq:cost} becomes $\mathcal{C}(\boldsymbol{\theta}) = 1 - P_0$. Its derivative then follows the rule
\begin{align}
\frac{\partial \mathcal{C}(\boldsymbol{\theta})}{\partial \theta_j}
%= \frac{-P_0(\boldsymbol{\theta} + \tfrac{\pi}{2}\mathbf{e}_j) + P_0(\boldsymbol{\theta} - \tfrac{\pi}{2}\mathbf{e}_j)}{2}.
= \dfrac{\mathcal{C}(\boldsymbol{\theta} + \tfrac{\pi}{2}\mathbf{e}_j)-\mathcal{C}(\boldsymbol{\theta} - \tfrac{\pi}{2}\mathbf{e}_j)}{2}.
\end{align}
Although this expression resembles the standard PSR, implementing it requires measurements on a 
$2n$-qubit circuit. Applying the standard PSR directly to Eq.~\eqref{eq:cost} would yield an incorrect gradient.

\begin{figure}[b]
    \centering 
\includegraphics[width=\columnwidth]{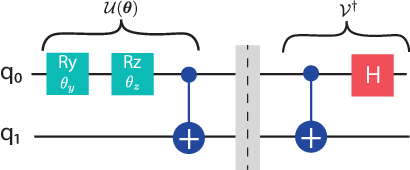} 
    \caption{Quantum circuit diagram: The circuit starts by applying \( \mathcal{U}(\bm\theta) \), followed by \( \mathcal{V}^\dagger \), and then measuring the final circuit.}
    \label{fig3} 
\end{figure}

\section{Demonstration}\label{sec4}
In this section, we consider three representative examples: (A) GHZ state preparation,  (B) Toffoli gate synthesis, and (C) random gate synthesis.
The GHZ example is used to illustrate our unitary PSR in a setting where the gradient expressions can be analyzed in detail and shown to agree exactly with theoretical predictions. The Toffoli example demonstrates the numerical performance of the method for a nontrivial unitary target, where we employ a hardware-efficient ansatz (HEA) and optimize its parameters to compile the target gate. Finally, the random gate synthesis example probes the behavior of the method in a more general setting, where the target unitary is drawn from the Haar measure.

\begin{figure*}
    \centering 
\includegraphics[width=0.7\linewidth]{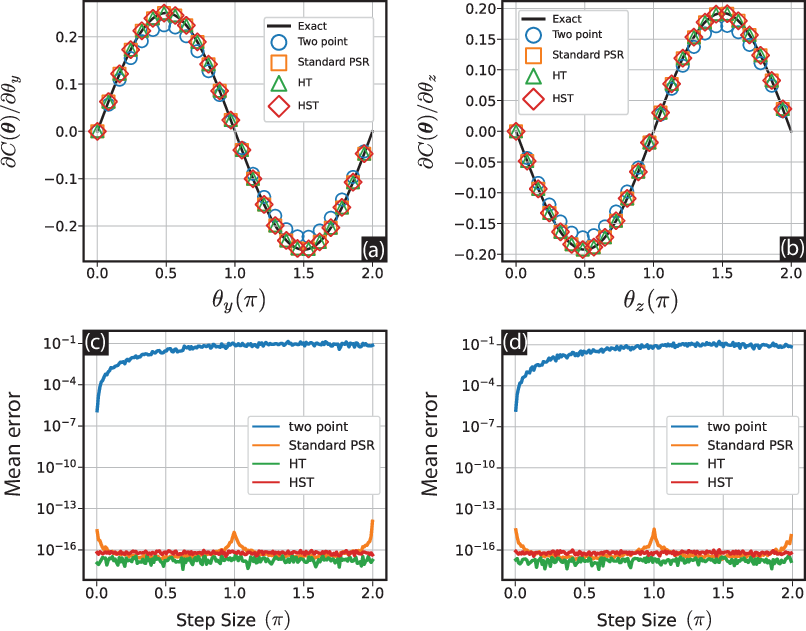} 
    \caption{Derivatives estimation results. (a, b) show the derivatives $\partial C(\theta)/\partial \theta_y$ and $\partial C(\theta)/\partial \theta_z$ as functions of $\theta_y$ and $\theta_z$, respectively. The unitary-based derivatives obtained using HT and HST are compared with the finite-difference two-point method, the standard PSR, and the exact analytical results. (c, d) present the average error of the derivative estimates as a function of the step size, highlighting the trade-off between accuracy and computational cost for the different methods.}
    \label{fig4} 
\end{figure*}

\subsection{GHZ state preparation}
To demonstrate our results, we consider a 2-qubit system with a target GHZ state. The operator \( \mathcal{U}(\bm\theta) \) is constructed by applying \( R_y(\theta_y) \) and \( R_z(\theta_z) \) gates to the first qubit, followed by a CNOT gate, where the first qubit serves as the control and the second qubit as the target. The operator \( \mathcal{V}^\dagger \) is formed by applying a Hadamard gate to the first qubit, followed by a CNOT gate with the same control-target configuration as in \( \mathcal{U}(\bm\theta) \). The quantum circuit is depicted in Fig.~\ref{fig3}. We begin by applying \( \mathcal{U}(\bm\theta) \) to the quantum circuit, followed by \( \mathcal{V}^\dagger \), and then measuring the final circuit.

To evaluate the gradient of the cost function \eqref{eq:cost}, we first analyze the composite operator $\mathcal{V}^\dagger \mathcal{U}(\bm\theta)$ as
\begin{align}
   \mathcal{V}^\dagger \mathcal{U}(\bm\theta) = W \cdot \Big(R_z(\theta_z) \otimes I\Big) \cdot \Big(R_y(\theta_y) \otimes I\Big),
\end{align}
where $W = \text{CNOT} \cdot (H \otimes I) \cdot \text{CNOT}$.

Using the unitary-based PSR Eq.~\eqref{eq:theory}, we have 
\begin{align}
\frac{\partial \mathcal{C}(\bm\theta)}{\partial \theta_y}
&= -\frac{1}{d^2}\,\text{Re} \!\Big[
\text{tr}\big(\mathcal{V}^\dagger \mathcal{U}(\theta_y+\pi, \theta_z)\big) \cdot 
\text{tr}\big(\mathcal{V}^\dagger \mathcal{U}(\theta_y, \theta_z)\big)^* \Big] \nonumber \\
&= \frac{4}{d^2} \, \sin(\theta_y) \, \sin^2\!\big(\tfrac{\theta_z}{2}\big),
\label{eq:dyC} \\[6pt]
\frac{\partial \mathcal{C}(\bm\theta)}{\partial \theta_z}
&= -\frac{1}{d^2}\,\text{Re} \!\Big[
\text{tr}\big(\mathcal{V}^\dagger \mathcal{U}(\theta_y, \theta_z+\pi)\big) \cdot 
\text{tr}\big(\mathcal{V}^\dagger \mathcal{U}(\theta_y, \theta_z)\big)^* \Big] \nonumber \\
&= -\frac{4}{d^2} \, \cos^2\!\Big(\tfrac{\theta_y}{2}\Big)\,\sin(\theta_z).
\label{eq:dzC}
\end{align}

Figure~\ref{fig4}(a, b) shows the partial derivative $\partial C(\bm\theta)/\partial \theta_y$ as a function of $\theta_y$ and $\partial C(\bm\theta)/\partial \theta_z$ as a function of $\theta_z$, respectively. In both cases, the derivatives obtained from the HT and HST match the theoretical analysis, confirming the accuracy of the unitary-based approach. In contrast, the finite-difference two-point method exhibits visible deviations from the exact gradients, which strongly depend on the chosen step size. The standard PSR performs significantly better than the finite-difference approach and captures the overall trend of the exact gradient. However, small systematic deviations remain compared to HT and HST. This comparison highlights that while the standard PSR improves upon classical numerical differentiation, the unitary-based methods provide higher accuracy and consistency.

To assess the precision and stability of these methods, we analyze the mean error of derivative estimation as a function of the step size. The step size is varied uniformly from 0 to $2\pi$, and for each value we evaluate the derivatives at $10^3$ randomly chosen parameter points sampled uniformly from the full space. The results are presented in Fig.~\ref{fig4}(c) for $\theta_y$ and Fig.~\ref{fig4}(d) for $\theta_z$.

As shown in Figs.~\ref{fig4}(c) and (d), the two-point finite-difference method is highly sensitive to the step size: very small steps yield accurate results but are impractical due to numerical instability, while larger steps introduce substantial bias. The standard PSR exhibits improved stability compared to finite differences, but its error remains higher than that of HT and HST at some points. In contrast, both HT and HST consistently achieve near machine-precision accuracy across the entire range of step sizes, demonstrating their numerical stability and reliability.

These results confirm that unitary-based gradient estimation methods such as HT and HST provide stable and precise gradients, outperforming both finite-difference schemes and the standard PSR. This robustness makes them particularly well suited for reliable gradient evaluation in VQAs.

\begin{figure}[t]
    \centering 
\includegraphics[width=\linewidth]{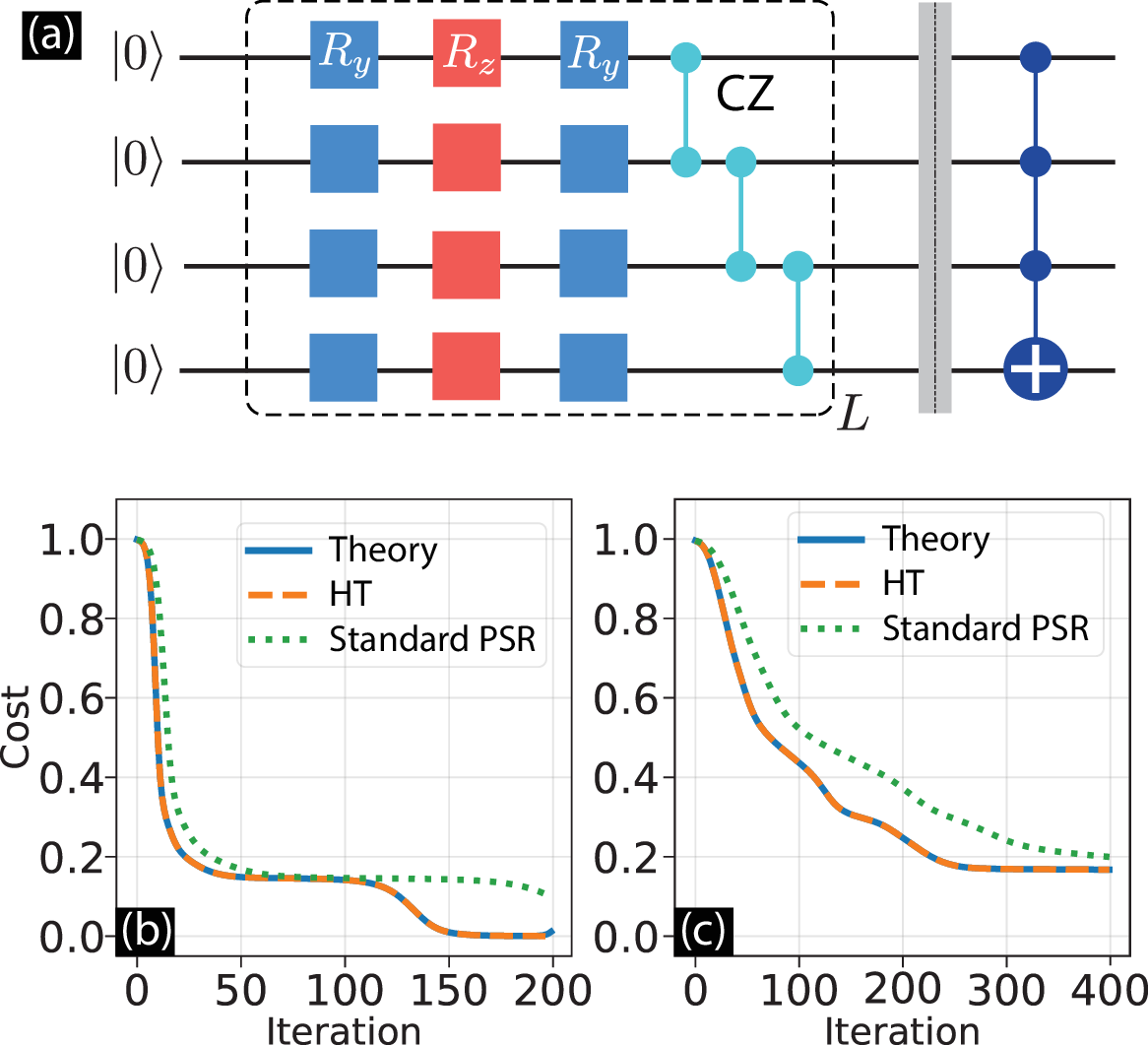} 
    \caption{(a) Quantum circuit used for Toffoli gate synthesis. (b, c) Cost versus iteration for 3-qubit and 4-qubit Toffoli gate synthesis, respectively. Gradient descent is used with learning rate \(1.5\) and HEA depth \(L=8\) and \(L=14\), respectively. The unitary-based PSR implemented through the HT matches the theoretical JAX gradients, whereas the standard PSR deviates from the theory. In the 4-qubit case, all methods converge to a local minimum due to the limited expressibility of the chosen ansatz.}
    \label{fig5} 
\end{figure}

\subsection{Toffoli gates synthesis}

In this subsection, we apply VQC to the synthesis of $N$-qubit Toffoli gates. The target unitary $\mathcal{V}$ is the standard Toffoli gates for $N=3$ and $N=4$. The compilation ansatz $\mathcal{U}(\bm{\theta})$ is implemented as a hardware-efficient ansatz (HEA), consisting of parameterized single-qubit rotations and nearest-neighbor entangling layers \cite{Kandala2017}. Each layer applies $R_y, R_z,$ and $R_z$ rotations to all qubits, followed by a chain of CZ gates that couple adjacent qubits. The quantum circuit is shown in Fig.~\ref{fig5} (a) for $N = 4$ case.
To optimize the parameters, we employ three approaches: a unitary-based PSR implemented through the HT, the standard PSR applied directly to the cost function in Eq.~\eqref{eq:cost}, and exact gradients computed using JAX-based automatic differentiation \cite{jax2018github}, which serve as the theoretical reference for benchmarking the numerical performance.

Figure~\ref{fig5}(b, c) shows the cost function versus the iteration for both system sizes. The HEA depth is set to $L=8$ for $N=3$ and $L=14$ for $N=4$, and a standard gradient-descent optimizer with the learning rate $1.5$ is used.
For the 3-qubit case in Fig.~\ref{fig5}(b), the unitary-based PSR reproduces the theoretical curve, while the standard PSR clearly deviates from the expected behavior. This matches the analysis in Sec.~\ref{sec3}, where applying the standard PSR directly to Eq.~\eqref{eq:cost} mixes the parameter shifts across the trace, producing biased gradients and leading to incorrect updates during optimization.

A similar trend appears in Fig.~\ref{fig5}(c) for the 4-qubit Toffoli gate. The unitary-based PSR follows the theoretical reference, while the standard PSR again deviates from the expected trajectory. In this case, all methods reach a local minimum, indicating the increased difficulty of compiling larger gates. 
The chosen HEA provides limited expressibility for $N=4$, resulting in a shallow cost landscape. Better performance could be achieved by increasing the circuit depth or adopting more expressive ans\"atze. However, such design choices are beyond the scope of this work, which aims to show that the unitary-based PSR consistently provides unbiased and stable gradients, whereas the standard PSR fails even in moderately challenging synthesis problems.

Similar gradient-evaluation procedures for specific circuit constructions were reported in Ref.~\cite{Khatri2019quantumassisted}.
These results are consistent with our formulation and can be obtained as special cases of our unitary-based approach, which derives the shift rule directly at the level of the parameterized unitary and applies to general differentiable circuit cost functions.

\subsection{Random unitary gate synthesis}
In this subsection, we study the general task of synthesizing random unitary gates.
The target unitary $\mathcal{V}$ is a Haar random, and the goal is to approximate
$\mathcal{V}^\dagger$ using a parameterized quantum circuit $\mathcal{U}(\boldsymbol{\theta})$.
As in the previous sections, we employ a HEA for $\mathcal{U}(\boldsymbol{\theta})$
and optimize its parameters via the VQC.

We compare three gradient-evaluation methods: the proposed HT method, the standard PSR, and the finite-difference scheme based on a two-point approximation.
Numerical simulations are performed for $N=2$ to $6$ qubits. For each system size,
the circuit depth is set to $5N$, and the optimization is carried out for $400$ iterations
using a learning rate of $1.5$.

\begin{figure}[t]
    \centering 
\includegraphics[width=\linewidth]{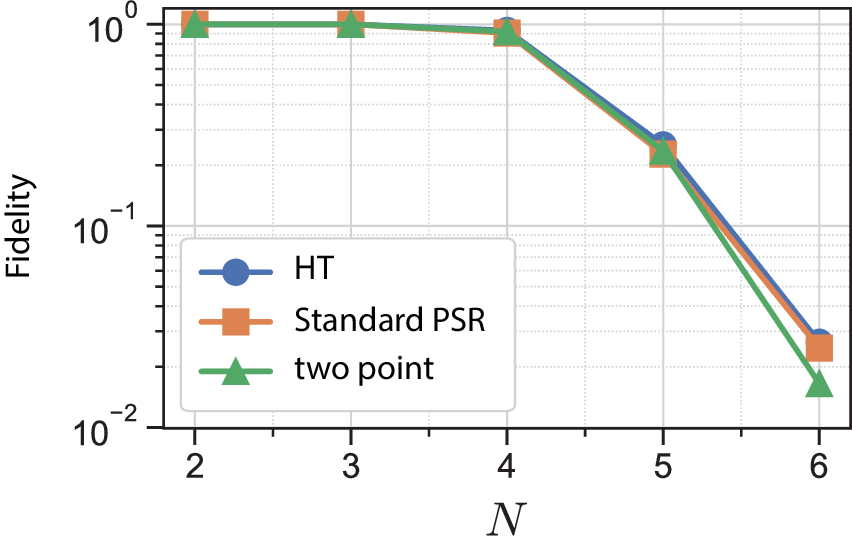} 
    \caption{Comparison of the fidelity for random unitary gate synthesis using
    the HT method, the standard PSR, and the two-point finite-difference
    approximation, as a function of the number of qubits $N$.}
    \label{fig6} 
\end{figure}

The resulting fidelities are shown in Fig.~\ref{fig6}. For small system sizes
($N \leq 4$), all methods achieve high fidelity and exhibit similar performance.
However, as $N$ increases, the optimization becomes more challenging and the overall
fidelity decreases. In this regime, noticeable deviations between the methods emerge.
In particular, the HT method shows slightly higher fidelities than the standard PSR and the two-point approximation in this regime.

\section{Discussion}\label{sec5}
The unitary-based PSR relaxes assumptions required by the standard rule and, as a result, can be applied to a broader range of VQA cost functions. It allows exact gradient evaluation even when the cost involves composite quantities or classical post-processing, such as those common in quantum machine learning and variational compilation.

A next step is to examine how this framework performs on NISQ hardware, where noise can directly influence the cost landscape. Incorporating noise-aware gradients and combining the method with advanced classical optimizers may further improve the stability of hybrid quantum-classical algorithms.

These developments also connect with emerging quantum-centric computing architectures \cite{ALEXEEV2024666,buchs2025rolequantumcomputingadvancing}, where quantum processors serve as specialized accelerators within larger HPC systems. In such settings, efficient and general gradient evaluation can help integrate quantum subroutines more effectively with classical optimization workflows.

\section{Conclusion}\label{sec6}
We have presented a unitary-based parameter-shift framework that provides exact gradients for any cost function that depends differentially on the parameters of a quantum circuit. This method works beyond standard expectation-value settings and can be used for tasks such as variational compilation, quantum machine learning, and other composite or nonlinear objectives. By offering a practical way to compute gradients, the approach improves the reliability of VQAs and may help support the development of more flexible quantum algorithms in computing, optimization, and metrology.

\begin{acknowledgments}
We would like to thank Vu Tuan Hai for valuable discussions. This paper is supported by JSPS
KAKENHI Grant Number 23K13025 and by the Tohoku Initiative for Fostering Global Researchers for Interdisciplinary Sciences (TI-FRIS) of MEXT's Strategic Professional Development Program for Young Researchers.
\end{acknowledgments}

\

\section*{Data availability} 
The data that support the findings of this article are openly
available \cite{urbaneja2025exactgrad}

\appendix

\section{Cost function derivation}
In this appendix, we first present the detailed calculations of Eqs.~(\ref{eq:dyC}, \ref{eq:dzC}) using the exact solution. We then prove that these results are also theoretical results by directly calculating the derivatives.

The operator \( \mathcal{V}^\dagger \)\( \mathcal{U} \) is explicitly given by
\begin{align}
   \mathcal{V}^\dagger \mathcal{U}(\bm\theta) = W \cdot \Big(R_z(\theta_z) \otimes I\Big) \cdot \Big(R_y(\theta_y) \otimes I\Big),
\end{align}
where 
\begin{align}
    \notag W = \text{CNOT} \cdot (H \otimes I) \cdot \text{CNOT} 
    = \frac{1}{\sqrt{2}}
    \begin{pmatrix}
    1 & 0 & 0 & 1 \\
    0 & 1 & 1 & 0 \\
    0 & 1 & -1 & 0 \\
    1 & 0 & 0 & -1
    \end{pmatrix},
\end{align}
$R_z(\theta_z) =
    \begin{pmatrix}
    e^{-i\theta_z/2} & 0 \\
    0 & e^{i\theta_z/2}
    \end{pmatrix}$, and 
    $R_y(\theta_y) = 
\begin{pmatrix}
\cos\!\big(\tfrac{\theta_y}{2}\big) & -\sin\!\big(\tfrac{\theta_y}{2}\big) \\
\sin\!\big(\tfrac{\theta_y}{2}\big) & \cos\!\big(\tfrac{\theta_y}{2}\big)
\end{pmatrix}$.

\subsection{Unitary-bases PSR solution}
To prove Eqs.~(\ref{eq:dyC}, \ref{eq:dzC}), we first derive 
$\operatorname{tr}(\mathcal{V}^\dagger \mathcal{U})^*$ straightforwardly as
\begin{align}\label{eq:trvutheta}
\operatorname{tr}\Big(\mathcal{V}^\dagger \mathcal{U}(\bm\theta)\Big)^*
= 2 \sqrt{2}\, i\, \sin\!\big(\tfrac{\theta_z}{2}\big) \cos(\tfrac{\theta_y}{2}).
\end{align}
Similarly, we have 
%To show this, we first evaluate the trace term $\text{tr}(\mathcal{V}^\dagger \mathcal{U}(\bm\theta+\pi\bm{e}_y))$ by shifting $\theta_y \to \theta_y + \pi$:
\begin{align}\label{eq:trvuthetapi}
\notag \text{tr}\Big(\mathcal{V}^\dagger \mathcal{U}(\theta_y+\pi, \theta_z)\Big) &= -2 \sqrt{2}\, i\, \sin(\tfrac{\theta_z}{2}) \cos(\tfrac{\theta_y+\pi}{2}) \\
 &= 2 \sqrt{2}\, i\, \sin(\tfrac{\theta_z}{2}) \sin(\tfrac{\theta_y}{2}).
\end{align} 
Substituting Eqs. (\ref{eq:trvutheta}, \ref{eq:trvuthetapi}) into Eq.~\eqref{eq:theory}, we have
\begin{align}\label{eq:AppdyC}
\frac{\partial \mathcal{C}(\bm\theta)}{\partial \theta_y}
&= -\frac{1}{d^2}\,\text{Re} \!\Big[
\text{tr}\big(\mathcal{V}^\dagger \mathcal{U}(\theta_y+\pi, \theta_z)\big) \cdot \text{tr}\big(\mathcal{V}^\dagger \mathcal{U}(\theta_y, \theta_z)\big)^* \Big] \nonumber \\
\notag&= -\frac{1}{d^2}\,\text{Re} \!\Bigg[
\left( 2 \sqrt{2}\, i\, \sin\!\big(\tfrac{\theta_z}{2}\big) \sin\!\big(\tfrac{\theta_y}{2}\big) \right) \cdot \\
\notag &\hspace{3cm}
\left( 2 \sqrt{2}\, i \,\sin\!\big(\tfrac{\theta_z}{2}\big) \cos\!\big(\tfrac{\theta_y}{2}\big) \right)
\Bigg] \nonumber \\
&= -\frac{1}{d^2}\,\text{Re} \!\Big[
(8) (i^2) \sin^2\!\big(\tfrac{\theta_z}{2}\big) \sin\!\big(\tfrac{\theta_y}{2}\big) \cos\!\big(\tfrac{\theta_y}{2}\big) \Big] \nonumber \\
&= -\frac{1}{d^2}\,\text{Re} \!\Big[
-8 \sin^2\!\big(\tfrac{\theta_z}{2}\big) \left(\tfrac{1}{2}\sin(\theta_y)\right) \Big] \nonumber \\
&= \frac{4}{d^2} \, \sin(\theta_y) \, \sin^2\!\big(\tfrac{\theta_z}{2}\big),
\end{align}
which is Eq.~\eqref{eq:dyC} in the main text.

Similarly, we have
\begin{align}
\notag \text{tr}(\mathcal{V}^\dagger \mathcal{U}(\theta_y, \theta_z+\pi)) &= -2 \sqrt{2}\, i\, \sin(\tfrac{\theta_z+\pi}{2}) \cos(\tfrac{\theta_y}{2}) \\
&= -2 \sqrt{2}\, i\, \cos(\tfrac{\theta_z}{2}) \cos(\tfrac{\theta_y}{2}).
\end{align}
Substituting this into Eq.~\eqref{eq:theory}, we have
\begin{align}\label{eq:AppdzC}
\frac{\partial \mathcal{C}(\bm\theta)}{\partial \theta_z}
&= -\frac{1}{d^2}\,\text{Re} \!\Big[
\text{tr}\big(\mathcal{V}^\dagger \mathcal{U}(\theta_y, \theta_z+\pi)\big) \cdot \text{tr}\big(\mathcal{V}^\dagger \mathcal{U}(\theta_y, \theta_z)\big)^* \Big] \nonumber \\
&= -\frac{1}{d^2}\,\text{Re} \!\Bigg[
\left( -2 \sqrt{2}\, i\, \cos\!\big(\tfrac{\theta_z}{2}\big) \cos\!\big(\tfrac{\theta_y}{2}\big) \right) \cdot \nonumber\\
&\hspace{3cm}
\left( 2 \sqrt{2}\, i \,\sin\!\big(\tfrac{\theta_z}{2}\big) \cos\!\big(\tfrac{\theta_y}{2}\big) \right)
\Bigg] \nonumber \\
&= -\frac{1}{d^2}\,\text{Re} \!\Big[
(-1)(8) (i^2) \cos^2\!\big(\tfrac{\theta_y}{2}\big) \sin\!\big(\tfrac{\theta_z}{2}\big) \cos\!\big(\tfrac{\theta_z}{2}\big) \Big] \nonumber \\
&= -\frac{1}{d^2}\,\text{Re} \!\Big[
8 \cos^2\!\big(\tfrac{\theta_y}{2}\big) \left(\tfrac{1}{2}\sin(\theta_z)\right) \Big] \nonumber \\
&= -\frac{4}{d^2} \, \cos^2\!\Big(\tfrac{\theta_y}{2}\Big)\,\sin(\theta_z).
\end{align}
which is Eq.~\eqref{eq:dzC} in the main text.

\subsection{Theoretical derivation}
The derivatives of the cost function can also be obtained through direct differentiation. In the following, we derive it explicitly and compare the outcome with the previous results. Taking the direct derivatives of the operator $\mathcal{V}^\dagger \mathcal{U}$ yields
\begin{align}
    \frac{\partial (\mathcal{V}^\dagger \mathcal{U})}{\partial \theta_y} 
    & = W \cdot R_z(\theta_z) \cdot \frac{\partial R_y(\theta_y)}{\partial \theta_y},
    \label{eq:part_1} \\
        \frac{\partial (\mathcal{V}^\dagger \mathcal{U})}{\partial \theta_z} 
    &= W \cdot \frac{\partial R_z(\theta_z)}{\partial \theta_z} \cdot R_y(\theta_y),
    \label{eq:part_2}
\end{align}
where we omitted $\otimes I$ for short.
The derivative of \(R_y(\theta_y)\) gives
\begin{equation}
    \frac{\partial R_y(\theta_y)}{\partial \theta_y} 
    = \tfrac{1}{2}
    \begin{pmatrix}
    -\sin(\tfrac{\theta_y}{2}) & -\cos(\tfrac{\theta_y}{2}) \\
    \cos(\tfrac{\theta_y}{2})  & -\sin(\tfrac{\theta_y}{2})
    \end{pmatrix}.
    \label{eq:part_4}
\end{equation}
Then, we have
\begin{align}
    R_z(\theta_z)\frac{\partial R_y(\theta_y)}{\partial \theta_y} 
    = \tfrac{1}{2}
    \begin{pmatrix}
    -e^{-i\tfrac{\theta_z}{2}}\sin(\tfrac{\theta_y}{2}) & -e^{-i\tfrac{\theta_z}{2}}\cos(\tfrac{\theta_y}{2}) \\
     e^{i\tfrac{\theta_z}{2}}\cos(\tfrac{\theta_y}{2}) & -e^{i\tfrac{\theta_z}{2}}\sin(\tfrac{\theta_y}{2})
    \end{pmatrix}.
    \label{eq:part_5}
\end{align}

Combining these components, the full derivative in \eqref{eq:part_1} can be expressed explicitly
\begin{widetext}
\begin{align}
    \frac{\partial (\mathcal{V}^\dagger \mathcal{U})}{\partial \theta_y}
    = \frac{1}{2\sqrt{2}}
    \begin{pmatrix}
    -e^{-i\theta_z/2}\sin\!\left(\tfrac{\theta_y}{2}\right) & e^{i\theta_z/2}\cos\!\left(\tfrac{\theta_y}{2}\right) & -e^{-i\theta_z/2}\cos\!\left(\tfrac{\theta_y}{2}\right) & -e^{i\theta_z/2}\sin\!\left(\tfrac{\theta_y}{2}\right) \\
    e^{i\theta_z/2}\cos\!\left(\tfrac{\theta_y}{2}\right)   & -e^{-i\theta_z/2}\sin\!\left(\tfrac{\theta_y}{2}\right) & -e^{i\theta_z/2}\sin\!\left(\tfrac{\theta_y}{2}\right) & -e^{-i\theta_z/2}\cos\!\left(\tfrac{\theta_y}{2}\right) \\
    -e^{i\theta_z/2}\cos\!\left(\tfrac{\theta_y}{2}\right)  & -e^{-i\theta_z/2}\sin\!\left(\tfrac{\theta_y}{2}\right) &  e^{i\theta_z/2}\sin\!\left(\tfrac{\theta_y}{2}\right) & -e^{-i\theta_z/2}\cos\!\left(\tfrac{\theta_y}{2}\right) \\
    -e^{-i\theta_z/2}\sin\!\left(\tfrac{\theta_y}{2}\right) & -e^{i\theta_z/2}\cos\!\left(\tfrac{\theta_y}{2}\right) & -e^{-i\theta_z/2}\cos\!\left(\tfrac{\theta_y}{2}\right) &  e^{i\theta_z/2}\sin\!\left(\tfrac{\theta_y}{2}\right)
    \end{pmatrix}.
\end{align}
\end{widetext}
The trace of this matrix gives the following.
\begin{align}
\text{tr}\!\left(\frac{\partial (\mathcal{V}^\dagger \mathcal{U})}{\partial \theta_y}\right)
= i \sqrt{2} \, \sin(\theta_y) \, \sin\!\big(\tfrac{\theta_z}{2}\big).
\label{eq:part7}
\end{align}
Finally, substituting Eqs.~(\ref{eq:trvutheta}, \ref{eq:part7}) into the cost function derivative (\ref{eq:exact}) gives
\begin{align}
\frac{\partial \mathcal{C}(\bm\theta)}{\partial \theta_y} 
= \frac{4}{d^2} \, \sin(\theta_y) \, \sin^2\!\big(\tfrac{\theta_z}{2}\big)
\label{eq:part_8_simplified}
\end{align}
This result matches the exact result in Eq.~(\ref{eq:AppdyC}). 

Similarly, the partial derivative with respect to $\theta_z$ in (\ref{eq:part_2}) gives
\begin{equation}
\frac{\partial (\mathcal{V}^\dagger \mathcal{U})}{\partial \theta_z}
 = W \cdot \Big(\tfrac{\partial R_z(\theta_z)}{\partial \theta_z}\, R_y(\theta_y)\Big).
 \label{eq:part_9}
\end{equation}
First, we find the derivative of the \(R_z(\theta_z)\) matrix:
\begin{equation}
\frac{\partial R_z(\theta_z)}{\partial \theta_z}
 = \tfrac{1}{2}
 \begin{pmatrix}
-i\,e^{-i\theta_z/2} & 0 \\[2pt]
0 & i\,e^{i\theta_z/2}
\end{pmatrix}.
\label{eq:part_10}
\end{equation}
Substitute it into Eq.~\eqref{eq:part_9}
and take the trace
\begin{equation}
\text{tr}\!\left(\frac{\partial (\mathcal{V}^\dagger \mathcal{U})}{\partial \theta_z}\right)
= -\,i\sqrt{2}\,\cos\!\Big(\tfrac{\theta_y}{2}\Big)\cos\!\Big(\tfrac{\theta_z}{2}\Big).
\label{eq:part_12}
\end{equation}
Finally, the cost function derivative gives the following
\begin{align}
\frac{\partial \mathcal{C}(\bm\theta)}{\partial \theta_z}
= -\frac{4}{d^2}\,\cos^2\!\Big(\tfrac{\theta_y}{2}\Big)\,\sin(\theta_z).
\label{eq:part_13}
\end{align}
This result matches the exact result in Eq.~(\ref{eq:AppdzC}).

\bibliography{ref}
\end{document}